\documentclass[twocolumn,aps,groupedaddress]{revtex4-2}

\usepackage{graphicx}
\usepackage{dcolumn}
\usepackage{bm}
\usepackage{amsmath,amssymb,amsfonts,MnSymbol}
\usepackage{float}
\usepackage{color}
\usepackage{dcolumn}
\usepackage{wasysym}
\usepackage{braket,mathtools}
\usepackage[normalem]{ulem}
\newcolumntype{N}{@{}m{0pt}@{}}

\begin{document}

\title{
Quantum anomalous Hall effect from inverted charge transfer gap
}

\author{Trithep Devakul}
\author{Liang Fu}
\affiliation{Department of Physics, Massachusetts Institute of Technology, Cambridge, Massachusetts 02139, USA}

\begin{abstract}
A general mechanism is presented by which topological physics arises in strongly correlated systems without flat bands. 
Starting from a charge transfer insulator, topology emerges when the charge transfer energy between the cation and anion is reduced to invert the lower Hubbard band and the spin-degenerate charge transfer band. 
A universal low-energy theory is developed for the inversion of charge transfer gap in a quantum antiferromagnet. 
The inverted state is found to be a quantum anomalous Hall (QAH) insulator 
with non-coplanar magnetism. 
Interactions play two essential roles in this mechanism: producing the insulating gap and quasiparticle bands prior to the band inversion, and causing the change of magnetic order necessary for the QAH effect after inversion.
Our theory explains the electric field induced transition from correlated insulator to QAH state in AB-stacked TMD bilayer MoTe$_2$/WSe$_2$. 
\end{abstract}

\maketitle


Electron correlation and band topology are two pivotal themes of quantum matter theory, which are deeply rooted 
in the particle and wave aspect of electrons respectively. 
The opposing traits of correlation and topology are clearly displayed in the contrast between a Mott insulator \cite{Mott} 
and a Chern insulator \cite{Haldane}. 
In a Mott insulator, electrons are bound to individual atoms and their motion is inhibited by mutual Coulomb repulsion. In contrast, a Chern insulator features chiral electrons on the boundary that refuse to localize.  
Despite their differences, electron correlation and band topology can cooperate to create 
fascinating states of matter, as exemplified by quantum Hall systems.
More recently, the scope of topology has been greatly expanded by 
the discovery of topological band insulators \cite{KaneHasan, QiZhang}
in numerous semiconductor materials \cite{FuKane2007,Zhang2019,Vergniory2019,Tang2019}.
Since then there has been great interest on  
interaction-induced topological states in correlated electron systems such as transition metal oxides 
and $f$-electron materials \cite{KimBalents, Fiete, Coleman, HuKane}. However, 
after considerable effort it remains unclear whether there is a {\it common} mechanism  for 
topological physics in {\it generic} Hubbard-type systems.

This work is an attempt to provide a guiding principle 
for the realization of strongly correlated topological states 
in materials with an odd number of electrons per unit cell. 
Building on and linking together the notions of Hubbard band, charge transfer gap, 
and topological band inversion, we find a simple and natural mechanism leading to 
magnetic topological insulators that exhibit 
{\it non-collinear} spin structures and quantum anomalous Hall effect. 

For systems with an odd number of electrons per unit cell, a large enough Coulomb repulsion $U$ can suppress double occupancy and produce an insulating state. 
In one-band Hubbard models at large $U$, the single-particle spectral function  consists of the lower and upper Hubbard bands separated by the Mott gap \cite{Hubbard}. 
More interesting and relevant to our work are charge-transfer insulators \cite{zaanen1985band, varma1987, zhang1988effective} (such as the cuprate).  
These materials are comprised of cations (Cu) and anions (O). Transferring an electron between the anion and the cation 
without creating double occupancy costs energy less than $U$. 
The physics of charge transfer insulators is captured by two-band Hubbard models, where the band derived from anions is located inside the Mott gap of the cation states. The insulating gap is thus controlled by the charge transfer energy $\Delta$ rather than $U$.     

The essence of our idea is that reducing the charge transfer energy $\Delta$ can induce band inversion 
between cation and anion Hubbard bands, and thereby drive a transition from a Mott insulator to a topologically nontrivial state in which cations and anions are highly entangled.   
The insulating state that emerges after this transition can viewed as having a {\it negative charge transfer gap}, in analogy with negative band gap in inverted semiconductors \cite{PbSnTe}. 


The band inversion paradigm 
is remarkably successful in understanding and predicting topological band insulators \cite{FuKane2007,BHZ, Zhang2009,Liu2008, Hsieh2012,Qian2014}. 
A prime example is SnTe, in which 
the cation Sn band and anion Te band are inverted around the $L$ points in the Brillouin zone \cite{Hsieh2012}.
The crucial difference here is that we consider correlated insulators with a {\it many-body gap} at half 
filling. The {\it quasiparticle bands} that emerge in such insulators 
generally bear no resemblance to the band structure in the noninteracting limit. 
Specifically, we show that the inversion of cation and anion Hubbard bands 
in an antiferromagnetic (AFM) charge transfer insulator 
leads to a Chern insulator with canted AFM order, as shown schematically in Fig.1. 
In order for this mechanism to work, certain conditions regarding 
the Hubbard band structure and the metal-ligand hybridization must be satisfied.

In bulk materials, the charge transfer energy $\Delta$ is largely determined by the chemistry of 
the underlying cation and anion. Recently, moir\'e superlattices in semiconductor heterostructures \cite{Wu2018,Wu2019, Tang2020, Regan2020, Shimazaki2020, Shabani2021} 
offer a physical realization of 
charge transfer insulators \cite{Zhang2020, Zhang2021, Slagle2020}, where $\Delta$ is highly tunable by the displacement field. Remarkably, a displacement-field-induced transition from a  correlated insulator to a quantum anomalous Hall (QAH) state has been discovered in AB-stacked MoTe$_2$/WSe$_2$ heterostructure \cite{Tingxin2021, ZhangPNAS}. 
However, the origin and nature of this QAH state is not understood.
Our theory explains the origin of this observed phenomena, and 
makes testable predictions.  

\begin{figure}[ht]
\includegraphics[width=\columnwidth]{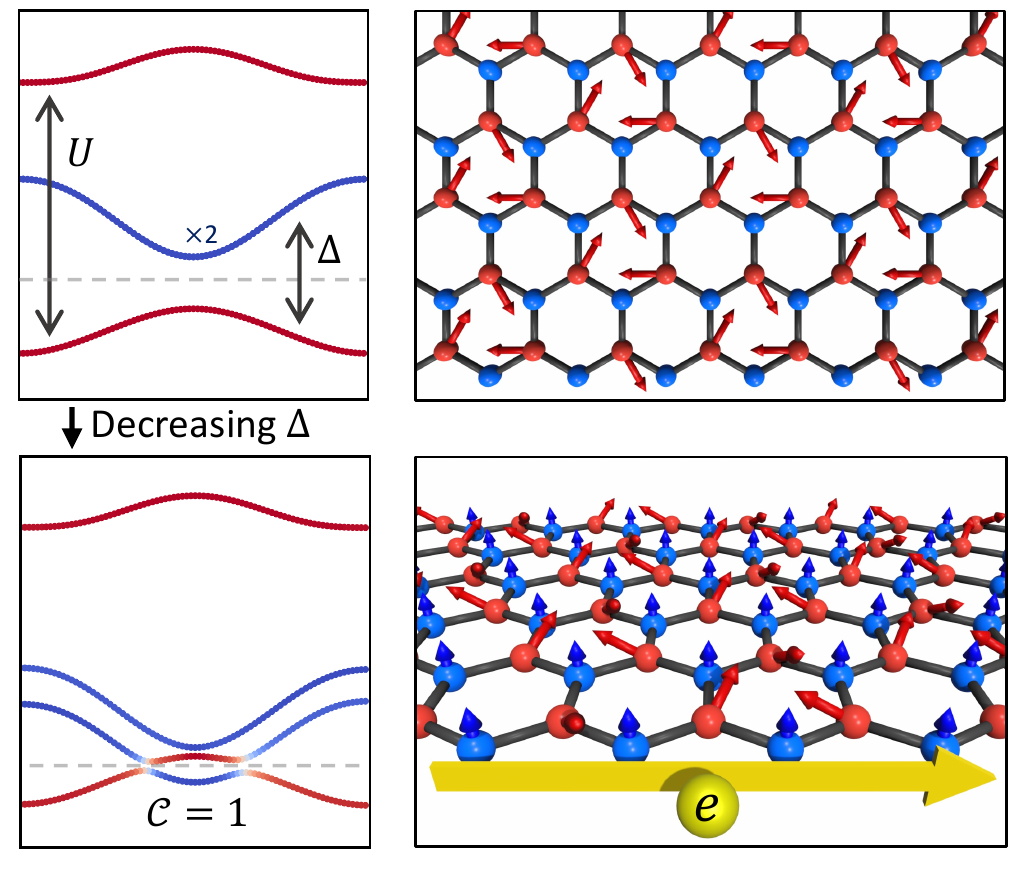}
\caption{
Our mechanism for topological Hubbard band inversion is illustrated.
(top) We start with a charge transfer insulator with $120^\circ$ $xy$ AFM order on the $A$ (red) sublattice.
The $A$ quasiparticle bands is split into a filled lower Hubbard band and an upper Hubbard band separated by energy $U$.
The spin-degenerate $B$ band lies in the Hubbard gap, resulting in a charge transfer insulator with charge transfer energy $\Delta$.
(bottom)
When $\Delta$ is reduced, a topological band inversion occurs. 
The Hubbard interaction on the $B$ (blue) sublattice results in spin splitting of the $B$ bands and non-coplanar spin order as illustrated.
The filled band has non-trivial Chern number and exhibits QAH.
}\label{fig:fig1}
\end{figure}

To illustrate the emergence of topology from inverted Hubbard bands, 
we introduce a two-band Hubbard model of spin-$\frac{1}{2}$ electrons on the honeycomb lattice, with the $A$ and $B$ sublattice 
representing the cation and anion respectively:   
\begin{equation}
\mathcal{H} = H_A + H_B + H_{AB} + U_A\sum_{i\in A} n_{i \uparrow} n_{i \downarrow} 
+ U_B\sum_{i\in B} n_{i \uparrow} n_{i \downarrow}. 
\end{equation}
Unless stated otherwise, we take $U_A=U_B= U$.
Here, $H_{A}$ and $H_B$ are the tight-binding Hamiltonian within each 
sublattice, while $H_{AB}$ is the hybridization term: 
\begin{eqnarray}
H_{\alpha} &=& - \sum_{\langle i,j\rangle_\alpha,\sigma} 
\left(t_\alpha  e^{i s_\sigma \nu_{ij}\phi_{\alpha}}  c^\dagger_{i\sigma} c_{j\sigma} + \rm{h.c.} \right)
-  \sum_{ i\in\alpha}  \frac{\tau_\alpha\Delta}{2} n_{i} \nonumber \\
H_{AB} &=& 
 -t_{AB}\sum_{\langle i,j\rangle,\sigma} c_{i\sigma}^\dagger c_{j\sigma}, 
\end{eqnarray}
where $\alpha = A, B$ denotes the two sublattices;  $\sigma = \uparrow, \downarrow$ electron spin $S^z$; $s_\uparrow=-s_\downarrow\equiv1$; $\tau_A= -\tau_B\equiv 1$. 
The first sum is over next nearest neighbor sites $\langle i,j\rangle_\alpha$, which belong to the same sublattice $\alpha$. Here, the hopping amplitude can be complex 
and spin-$s_z$ dependent, describing  
 an Ising spin-orbit coupling that is allowed by symmetry, 
with $\nu_{ij}=-\nu_{ji}=\pm 1$ depending on the path connecting site $j$ to $i$~\cite{Kane2005}. 
If the path turns right, $\nu_{ij}=1$, otherwise $\nu_{ij}=-1$. 
The special case $\phi_A = \phi_B = \frac{\pi}{2}$ corresponds to the 
original Kane-Mele model. 
The second term is a sublattice potential with $\Delta$ the charge transfer energy.
The last term is a hybridization term which connects nearest neighbor sites $\langle i,j\rangle$. 

Our model is motivated by and captures the essential physics of $\pm K$-valley spin-polarized moir\'e bands in transition metal dichalcogenides (TMD) bilayers. Consider for example MoTe$_2$/WSe$_2$.  
Its low-energy electron states  reside primarily in MM region of the MoTe$_2$ layer and XX region of the WSe$_2$ layer respectively, which together form a honeycomb lattice (Fig~\ref{fig:TMDFig}b)~\cite{ZhangPNAS}. The corresponding moir\'e bands are well described by our tight-binding Hamiltonian $H=H_A + H_B + H_{AB}$, with the sublattice and spin $\sigma$ corresponding to the layer and the $\pm K$ valley respectively. The charge transfer energy $\Delta$ corresponds to the layer bias potential, which is tuned by an applied displacement field. As we show in Supplementary Material, 
the relevant model parameters 
are $\phi_{A}\approx0$ and $\phi_B\approx-\frac{2\pi}{3}$.  
We shall focus on the case of $\phi_A=0$ for our following analysis.

At $\Delta\rightarrow \infty$, our model effectively reduces to the standard 
one-band Hubbard model on the triangular lattice of $A$ sites.   
As is well known, at the filling $n=1$ considered throughout this work, the ground state at large $U$ is a $120^\circ$ AFM ordered Mott insulator.
For large (but finite) $\Delta$, integrating out the $B$ sublattice gives an effective $A$ sublattice Hubbard model with complex $s_z$-dependent hopping parameters. Consequently, the effective spin model derived from our Hubbard model at large $U$ is an XXZ Heisenberg model with Dzyaloshinskii-Moriya interaction. The reduced spin $U(1)$ symmetry results in the $120^\circ$ AFM within the $xy$ plane, 
as shown in  Ref.~\cite{devakul2021magic}. 
This AFM insulator serves as the starting point of our analysis below.   


In order to obtain the complete phase diagram of our model, we 
perform a Hartree-Fock (HF) treatment of the Hubbard interaction.
We expect HF to give qualitatively reasonable results for insulating phases and, more importantly, 
will allow for an analytical understanding of the key physics through the quasiparticle band structure.
Our main findings from the HF study are fully supported by density matrix renormalization group (DMRG) calculations to be presented below.  

The HF approximation amounts to the replacement of the Hubbard term by
\begin{equation}
H_{\rm{Hub}}^{\rm{HF}} = \frac{U}{2} \sum_{i} \left(n_{i} \langle n_{i} \rangle - \vec{s}_{i}\cdot \langle\vec{s}_{i}\rangle -\frac{1}{2}\langle n_{i}\rangle^2 + \frac{1}{2}|\langle\vec{s}_{i}\rangle|^2\right)
\end{equation}
where $n_{i}=\sum_{\sigma}n_{i \sigma}$, $s^k_{i} = \langle c^\dagger_{i\sigma} \bm{\pi}^k_{\sigma\sigma^\prime} c_{i\sigma^\prime}\rangle$, and $\bm{\pi}^k$ are $k=x,y,z$ Pauli matrices.
The HF Hamiltonian must be solved self-consistently at the filling $n=1$ to 
obtain the HF ground state (in cases where multiple self-consistent solutions are found, the lowest energy solution among them is chosen).

In order to characterize the magnetic order, we define the following order parameter, a matrix in spin space:   
\begin{eqnarray}
S_{\sigma^\prime \sigma}^{\alpha\zeta} \equiv \frac{1}{N} \sum_{\bm{k}}\langle c^\dagger_{\alpha\sigma^\prime(\bm{k}+\zeta\bm{K})} c_{\alpha\sigma \bm{k}}\rangle 
\end{eqnarray}
where $\zeta=0$ describes ferromagnetic (FM) states, and $\zeta=\pm 1$ describes antiferromagnetic (AFM) states with wavevector $\pm \bm{K}=\pm \frac{4\pi}{3a}(1,0)$. 
In the presence of spin-orbit interaction, $xy$ AFM states with $\zeta=+$ and $-$ are distinct states displaying spin configurations of opposite chirality and are {\it not} degenerate~\cite{Zang2021}. For $\phi_A=0$ and $\phi_B\approx-\frac{2\pi}{3}$, 
the AFM Mott insulator at large $\Delta$ has $\zeta=-1$.

\begin{figure}[t]
\includegraphics[width=\columnwidth]{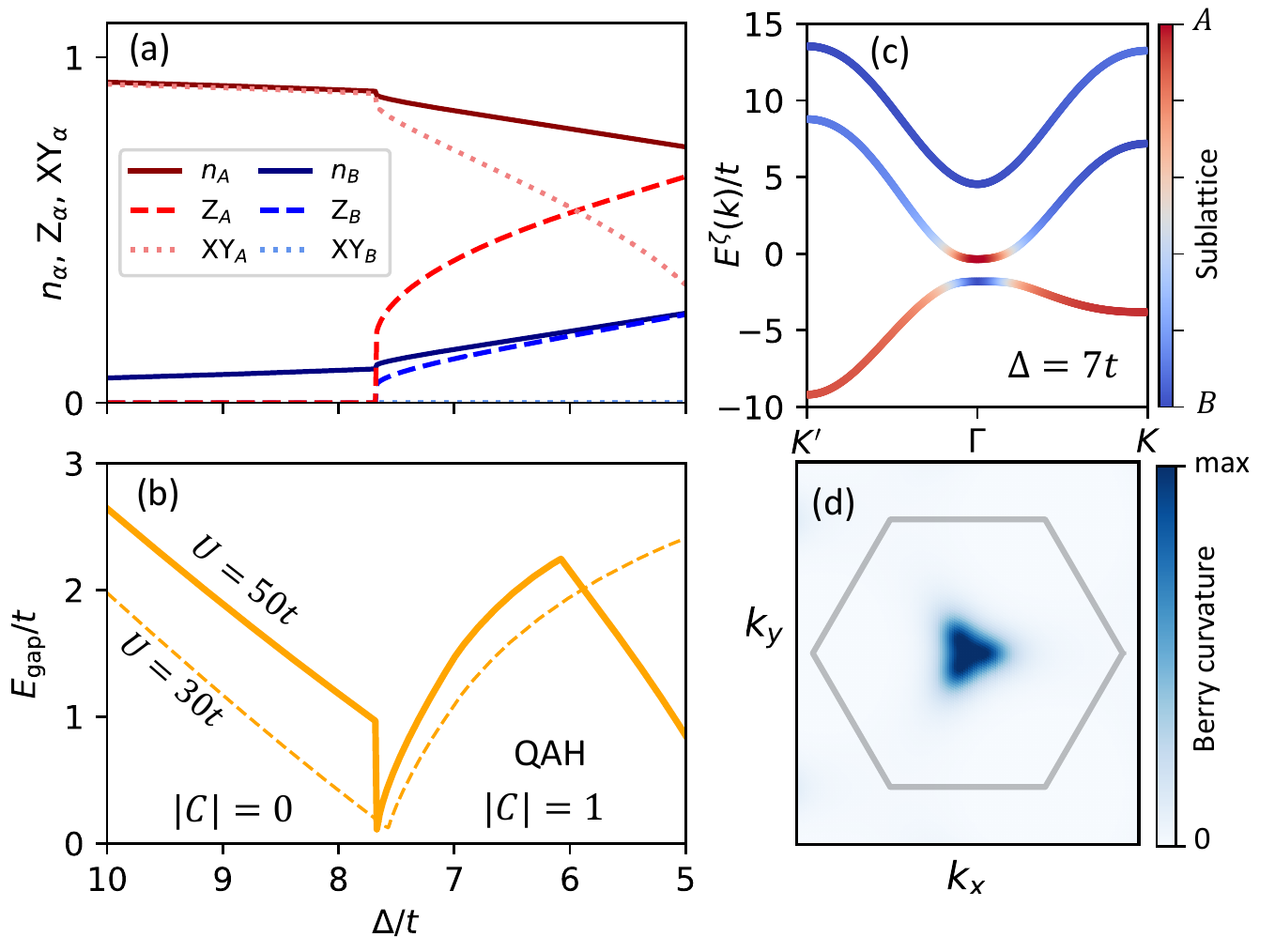}
\caption{
Order parameters (a) and charge gap (b) obtained from self-consistent HF as a function of $\Delta$ is shown, for parameters $t_A=t_B=\frac{1}{2}t_{AB}=t$, $\phi_A=0$, $\phi_B=-\frac{2\pi}{3}$, and $U=50t$.
There is a transition from the $xy$ AFM phase to the non-coplanar QAH phase with Chern number $|\mathcal{C}|=1$ as $\Delta$ is reduced.
The gap is also shown for $U=30t$.
(c) The quasiparticle band structure, with shift $\zeta=-$, obtained from HF at $\Delta=7t$ is shown.
The Berry curvature of the filled topological band is shown in (d).
}\label{fig:HFFig}
\end{figure}

The order parameters are shown in Fig~\ref{fig:HFFig}a as a function of $\Delta$ near the band inversion point for $\phi_B=-\frac{2\pi}{3}$, $t_A=t_B=\frac{1}{2}t_{AB} = t$, and $U=50t$.
We have defined the combinations
$\mathrm{Z}_{\alpha}\equiv \frac{1}{2}(S^{\alpha 0}_{\uparrow\uparrow} - S^{\alpha 0}_{\downarrow\downarrow})$ and $\mathrm{XY}_{\alpha} \equiv |S^{\alpha(\zeta=-1)}_{\uparrow\downarrow}|$, which capture the observed non-zero $z$ FM and $xy$ AFM orders respectively on the $\alpha$ sublattice.
Figure~\ref{fig:HFFig}b also shows charge gap and Chern number of the HF ground state.
We observe two distinct insulating phases:
a $xy$ AFM (with $\zeta=-$) on the $A$ sublattice at large $\Delta$ transitions into a canted $xy$ AFM as $\Delta$ is decreased.
This canted phase, in particular, has non-trivial Chern number $|\mathcal{C}|=1$, and is therefore a QAH phase.
This QAH phase with non-coplanar magnetism appearing at reduced charge transfer energy $\Delta$
is the highlight of this work.

To gain insight into the origin of the QAH phase, we examine the 
evolution of the quasiparticle band structure as a function of $\Delta$. As a first step, it is useful to 
first derive the noninteracting band structure at $U=0$. 
By Fourier transform, the single-particle Hamiltonian $\mathcal{H}_0=H_A+H_B+ H_{AB}$ takes the form
 $\mathcal{H}_0 = \sum_{\bm{k}\sigma} \vec{c}^\dagger_{\sigma\bm{k}} H_{\sigma}(\bm{k})\vec{c}_{\sigma\bm{k}}$, where $\vec{c}_{\sigma\bm{k}}^\dagger = (c_{A\sigma \bm{k}}^\dagger,c_{B\sigma \bm{k}}^\dagger)^T$ in $\bm k$ space, and
the Bloch Hamiltonian is
\begin{equation}
H_{\sigma}(\bm{k}) = 
\begin{pmatrix}
\mathcal{E}_{A\sigma}(\bm{k}) & T_{\sigma}(\bm{k})\\
T_{\sigma}^\dagger(\bm{k}) & \mathcal{E}_{B \sigma}(\bm{k}) 
\end{pmatrix}
\end{equation}
where
\begin{eqnarray}
\mathcal{E}_{\alpha\sigma}(\bm{k}) =& -2t_\alpha\sum_{n}\cos(\bm{k}\cdot \bm{a}_n + s_\sigma \tau_\alpha \phi_\alpha) - \frac{1}{2}\tau_\alpha\Delta \label{eq:singleparticleband}\\
T_{\sigma}(\bm{k}) =& -t_{AB}(e^{-i \bm{k}\cdot\bm{b}_1} + e^{-i \bm{k}\cdot\bm{b}_2} +  e^{-i\bm{k}\cdot\bm{b}_3}),
\end{eqnarray}
where $\bm{a}_n = a\left[\cos\frac{2\pi n}{3},\sin\frac{2\pi n}{3}\right]$ and $\bm{b}_n = \frac{a}{\sqrt{3}} \left[ \sin\frac{2\pi n}{3}, -\cos\frac{2\pi n}{3}\right]$.

At large $U$, the quasiparticle band structure of the $xy$ or canted AFM insulator is completely different from the noninteracting case.  
While the AFM order results in a $\sqrt{3}\times \sqrt{3}$ enlarged unit cell,  
this state is invariant under a combination of the unit translation and spin rotation around $z$ axis. 
Thanks to this symmetry property, the description of quasiparticle band structures can be 
simplified by performing a spin-dependent momentum boost with a unitary transformation $ U_\zeta : c^\dagger_{\uparrow \bm{k}} \rightarrow c^\dagger_{\uparrow (\bm{k}+\zeta \bm{K})},  
c^\dagger_{\downarrow \bm{k}} \rightarrow c^\dagger_{\downarrow (\bm{k} -\zeta \bm{K})}$. 
This transformation preserves the $z$ FM order 
and maps the $xy$ AFM order into $xy$ FM order, which is translationally invariant. 
After this transformation, the HF Hamiltonian, which includes the effect of magnetic order, is 
a $4 \times 4$ matrix (involving sublattice and spin) given by
\begin{equation}
H^{\rm{HF}}_\zeta(\bm{k}) = 
\begin{pmatrix}
 H_{\uparrow}(\bm{k}-\zeta\bm{K}) + \frac{U}{2}\mathrm{S}^{0}_{\downarrow\downarrow} & 
 -\frac{1}{2} U \rm{S}^\zeta_{\uparrow\downarrow} \\
 -\frac{1}{2} U (\rm{S}^{\zeta}_{\uparrow\downarrow})^*  &
 H_{\downarrow}(\bm{k}+\zeta\bm{K}) + \frac{U}{2}\rm{S}^{0}_{\uparrow\uparrow} 
\end{pmatrix}
\end{equation}
where $\mathrm{S}^{\zeta}_{\sigma^\prime\sigma} = \mathrm{diag}(S^{A\zeta}_{\sigma^\prime\sigma},S^{B\zeta}_{\sigma^\prime\sigma})$.

In the limit $\Delta \rightarrow \infty$, 
the two sublattices are decoupled and only the A  
sublattice is occupied at the filling of $n=1$, thus realizing the triangular lattice Hubbard model. 
In the $xy$ AFM insulator, the half-filled band splits into lower and upper Hubbard bands $E_\pm^{\zeta}(\bm{k})$, separated with the Mott gap $U$. 
In the large-$U$ limit, the lower Hubbard band associated with hole excitations 
has the energy dispersion
\begin{eqnarray}
E_{-}^{\zeta}(\bm{k}) = \frac{1}{2}[\mathcal{E}_{A\uparrow}(\bm{k}-\zeta\bm{K}) + \mathcal{E}_{A\downarrow}(\bm{k}+\zeta\bm{K})].   \label{hole}
\end{eqnarray}
Since the hopping amplitude of holes between adjacent sites 
is effectively reduced by the {\it noncollinear} spin configuration, 
 the bandwidth of holes is smaller than the noninteracting band, 
 but remains finite even as $U\rightarrow \infty$. This hole dispersion $E_-^{\zeta}(\bm{k})$ has a single maximum at $\Gamma$, which should be contrasted with the noninteracting band, $\mathcal{E}_{A\sigma}(\bm{k})$, which has two maxima. 

As the charge transfer energy $\Delta$ decreases below $U$, the $B$ sublattice band 
lies below the upper Hubbard band on the $A$ sublattice. This leads to  
a charge transfer insulator, in which low-energy hole and electron states reside primarily on  
 $A$ and $B$ sublattice respectively. 
While the hole band has a unique maximum at $\Gamma$ (after performing the transformation $U_\zeta$),   
the location of electron band minimum depends on the spin-orbit coupling parameter $\phi_B$.    
For $\frac{\pi}{3}<\zeta\phi_B<\pi$, there exist two degenerate minima: a $\sigma=\uparrow$ state at $\zeta K$ and $\downarrow$ at $-\zeta K$, both of which are shifted by the transformation $U_\zeta$ 
to $\Gamma$, coinciding with the hole band maximum. 
In such case, the charge transfer insulator has a direct gap. We then ask the question: 
what happens if $\Delta$ is decreased further so as to invert the charge transfer gap?    
 
To address this question, we develop a low-energy theory of 
hole and electron bands around $\Gamma$ near the gap inversion. 
Prior to the gap inversion, 
the $B$ sublattice is largely unoccupied, hence the
electron band is spin-degenerate. In contrast, due to 
the $xy$ AFM order, 
the lower Hubbard band associated with holes on the $A$ sublattice 
is spin-nondegenerate and
comprised of a superposition of $\sigma=\uparrow,\downarrow$ states.
The two bands are coupled by the hybridization term $H_{AB}$,  
which takes $p$-wave form near the gap. 
Taking $\zeta=-1$ and $\phi_B=-\frac{2\pi}{3}$ as in Fig~\ref{fig:HFFig}, we have
\begin{equation}
\begin{split}
T_{\sigma}(\bm{k} + s_\sigma\bm{K}) &\approx \frac{\sqrt{3}}{2}t_{AB} a s_\sigma (k_x - i s_\sigma k_y)  \\
&\equiv \sqrt{2} s_\sigma \lambda k_{s_\sigma},
\end{split}
\end{equation}
where $k_\pm\equiv k_x\pm i k_y$.   
By projecting the HF Hamiltonian into this low-energy subspace, 
we obtain a $k \cdot p$ theory of quasiparticle band structure in the $xy$ AFM 
state prior to gap inversion:  
\begin{equation}
H^{\rm{eff}}(\bm{k}) = 
\begin{pmatrix}
-\frac{\bm{k}^2}{2 m_A}  & \lambda k_- & -\lambda e^{i\theta} k_{+} \\
\lambda k_{+} & \frac{\bm{k}^2}{2m_B} + \delta & 0 \\
-\lambda e^{-i\theta} k_{-} & 0 & \frac{\bm{k}^2}{2 m_B} + \delta
\end{pmatrix} \label{eq:Heff}
\end{equation}
with $m_A = \frac{2}{3 t_A a^2}$ in the large $U$ limit, $m_B = \frac{1}{3 t_B a^2}$, $e^{i\theta}$ reflects the direction of in-plane order on the $A$ sublattice,
 and 
$\delta$ defines the charge transfer gap.

As the charge transfer gap $\delta$ is decreased and eventually becomes negative (while the charge transfer energy $\Delta$ remains positive), the occupation of $B$ sublattices increases, hence the effect of Hubbard repulsion $U_B$ between electrons becomes important. The low-energy theory of our charge transfer insulator, including one-particle term and two-body interaction, is: 
\begin{eqnarray}
\mathcal{H}^{\rm{eff}} = \sum_{\bm k} f^\dagger_{{\bm k} i } H_{ij}^{\rm{eff}}(\bm{k}) f_{{\bm k} j} + g \int d{\bm r}  \; 
n_{B\uparrow}({\bm r}) n_{B\downarrow}({\bm r}) \label{eq:effective}
\end{eqnarray}
where $f = (f_A, f_{B\uparrow}, f_{B\downarrow})$ denotes fermion quasiparticles;    
$n_{B\sigma} = f^\dagger_{B\sigma} f_{B\sigma}$, and 
the contact interaction $g$ is proportional to $U_B$. 
An additional interaction term $n_A n_B$ appears in the effective Hamiltonian $H$ 
when we further include nearest-neighbor interaction between $A$ and $B$ sites.  
Our {\it interacting} Hamiltonian $H$ captures the {\it universal} aspects of ``Hubbard band inversion'' in charge transfer insulators, in the same spirit as the Dirac Hamiltonian encapsulates band inversion in narrow gap semiconductors. 
However, there are fundamental differences between the two theories.
A charge transfer insulator has an inherent particle-hole asymmetry: holes associated with the lower Hubbard band are spin-nondegenerate, while electrons associated with the charge transfer band are spin-degenerate prior to the inversion. 
As a result, new physics arises after inverting the charge transfer gap, as show below.

We first analyze the quasiparticle energy spectrum at $g=0$, given by $H^{\rm{eff}}(\bm k)$.  
At ${\bm k}=0$ where the hybridization term vanishes,  
the spectrum consists of the spin-non-degenerate Hubbard band from the $A$ sublattice and the spin-degenerate band from the $B$ sublattice. 
Importantly, the two-fold degeneracy of the latter is protected by two symmetries of the $xy$ AFM state: 
(1) three-fold rotation of the lattice and electron spin around a hexagon center ($C_3$); (2) time-reversal transformation combined with a 
$\pi$ rotation of spin around $z$ axis ($i s_z \Theta$). Note that $(i s_z) \Theta$ is an anti-unitary 
symmetry operator that squares to identity, effectively acting as a time reversal operator in spinless systems. 
Thus,   the $B$ band at ${\bm k}=0$ furnishes a {\it real two-dimensional} representation of $C_3$. 

\begin{figure}[t]
\includegraphics[width=\columnwidth]{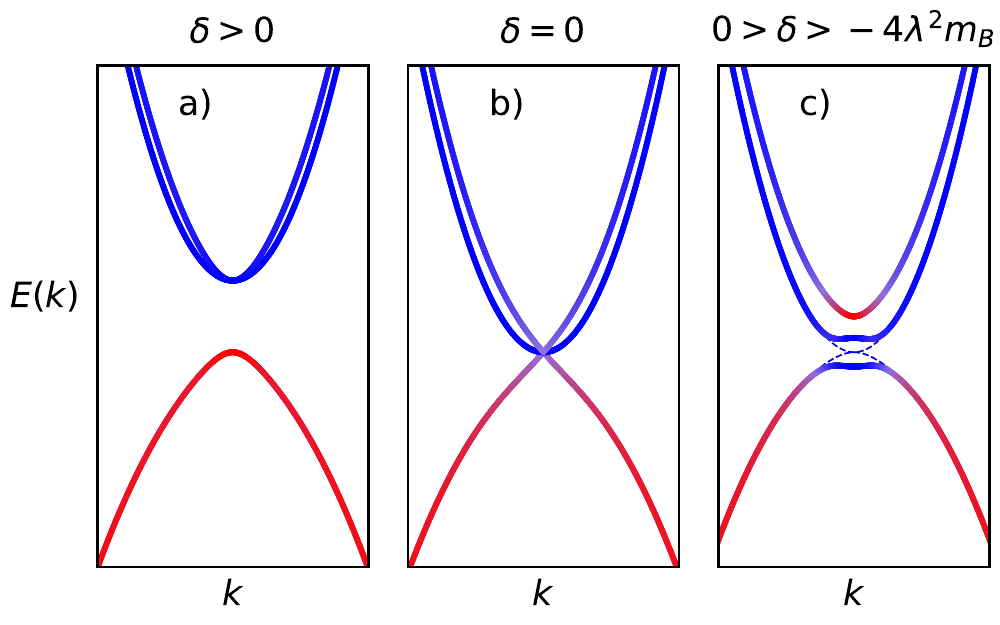}
\caption{
The band structure of the effective theory (Eq~\ref{eq:effective}) near inversion.
The band colors indicate the sublattice content, blue for $A$ and red for $B$ bands.
$(a)$ and $(b)$ show the bands before and at inversion.  
After inversion, $(c)$, the $g=0$ bands feature a quadratic band touching (dashed lines).
A perturbative instability then opens a topological gap for $g>0$, as illustrated.
For $\delta<-4\lambda^2m_B$, not shown, Fermi surfaces form and the system is metallic at $g=0$.
}\label{fig:FieldTheoryFig}
\end{figure}

Prior to gap inversion ($\delta>0$), the $B$ band lies above the $A$ band, and  
the Fermi level is inside the gap (Fig~\ref{fig:FieldTheoryFig}a). 
 We remark that at precisely $\delta=0$, the spectrum of $H^{\mathrm{eff}}$ consists a linearly dispersing Dirac cone and a parabolic electron band (Fig~\ref{fig:FieldTheoryFig}b).  
This critical point has no divergent susceptibility and we therefore expect it to be perturbatively stable to interactions.
When $\delta$ is tuned to become negative, 
the $B$ band dips below the $A$ band around ${\bm k}=0$. 
Due to the two-fold degeneracy of $B$ band at ${\bm k}=0$, 
the spectrum of $H^{\rm eff}$ immediately after band inversion, in the parameter range   
$-4 \lambda^2 m_B <\delta<0$, shows a quadratic band touching at the Fermi level (dashed line in Fig~\ref{fig:FieldTheoryFig}c), resulting in finite density of states for both electrons and holes. However, as shown by Sun, Yao, Fradkin and Kivelson \cite{Sun2009}, 
this kind of zero-gap state is unstable towards exciton condensation in the presence of even {\it arbitrarily weak} repulsive interaction. 
The interaction $g\propto U_B$ on the anions thus plays an essential role 
 after the charge transfer gap is inverted. 
The leading susceptibility of such a quadratic band touching is towards the opening of a \emph{topological} gap (solid lines in Fig~\ref{fig:FieldTheoryFig}c) resulting in a QAH state with spontaneous $Z_B\neq0$.
This analysis, based on the effective field theory Eq~\ref{eq:effective}, is controlled in the limit of small $U_B$.  

 Our HF calculation confirms the field theory analysis even beyond the small $U_B$ limit. 
Additionally, due to the $A-B$ hybridization, a finite occupation of $B$ sublattice is already present at $\delta>0$. 
This causes an upward shift in the energy of  the charge transfer band by $\frac{U_B}{2}\langle n_B\rangle$, which has the effect of delaying 
the transition to the inverted phase from $\delta=0$ to $\delta_c<0$.  
More importantly, in the presence of the Hubbard interaction $U_B$, a 
spontaneous spin polarization in the $\pm z$ direction is found at $\delta<\delta_c$,   
resulting in a non-coplanar spin structure with canted AFM on the $A$ sublattice and $z$ FM on the $B$ sublattice, as shown in Fig.1.  

In the noncoplanar phase, the $z$ FM order parameter component breaks the effective time-reversal symmetry $i s_z \Theta$, and produces spin splitting of the $B$ band. 
One of the spin-split bands is pushed to higher energy, while the other one takes part in the band inversion with the $A$ Hubbard band.  
Shown in Fig~\ref{fig:HFFig}d is the $\bm k$-space Berry curvature of the noncoplanar phase, obtained from the self-consistent HF Hamiltonian 
that includes both $xy$ AFM and $z$ FM orders. 
Now, the inversion around $\Gamma$ between $A$ and $B$ Hubbard bands---with {\it removed spin degeneracy}
and $p$-wave hybridization---gives rise to a QAH insulator with the Chern number $\mathcal{C}=\pm 1$ as computed directly from the Berry curvature integration. 

It is important to note that the appearance of QAH phase requires that the cation and anion Hubbard bands are dispersive, so that they can be inverted in a {\it part} of momentum space near the gap edge {\it before} $\Delta$ decreases to zero. 
 This is satisfied in our model since magnetic frustration of the cations leads to dispersive quasiparticle bands even for large $U$ (Eq~\ref{hole}). 
As such, the QAH phase is a consequence of the balance and synergy between electron localization and itinerancy.

\begin{figure}[t]
\includegraphics[width=\columnwidth]{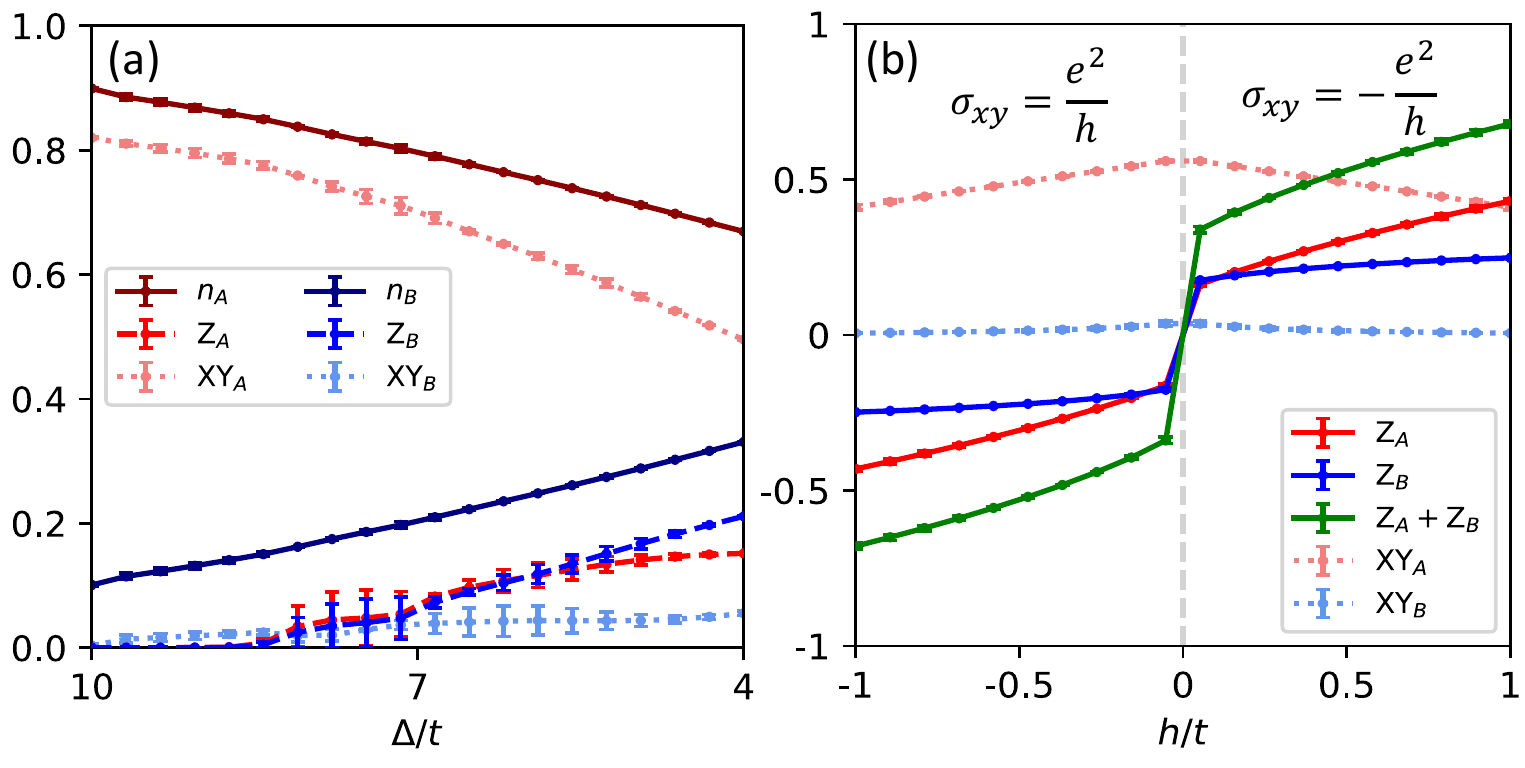}
\caption{
(a) Order parameters obtained from DMRG as a function of $\Delta$, showing qualitatively similar results as self-consistent HF.
The small $XY_B\neq0$ is likely due to the cylindrical geometry~\cite{supp}.
(b) The response to an applied Zeeman field $h$.
There is a discontinuity at $h=0$ due to broken symmetry.
While $Z_B$ is quickly saturated, $Z_A$ continues to increase with $h$ while $XY_A$ decreases, indicating a smooth variation in the canting.
The Hall conductivity $\sigma_{xy}$ changes sign discontinuously at $h=0$.
 Hysteresis is absent as we obtain the ground state independently for each value of $h$.
}\label{fig:DMRGFig}
\end{figure}


Our finding of the QAH phase with negative charge transfer gap and non-coplanar magnetism    
is further confirmed by DMRG calculations~\cite{White1992,Ostlund1995}.
Using the infinite DMRG algorithm, we study the ground state of the Hamiltonian on an infinite cylinder $L_x=\infty$ of circumference $L_y=6$ unit cells.
The unit cell in $x$ is chosen to be commensurate with the $\sqrt{3}\times\sqrt{3}$ AFM order.
More details on the numerical simulations and convergence of DMRG, performed using the TenPy code~\cite{tenpy}, is provided in the supplemental information.

Figure~\ref{fig:DMRGFig}a shows the order parameters as a function of $\Delta$, for the same set of parameters as before.
For each $\Delta$, we perform calculations for both periodic and anti-periodic boundary conditions in the 
circumferential direction. The difference in calculated observables, represented by the error bars, 
serves as an indication of finite-size effect~\cite{supp}.   
For a range around $\Delta\approx5t$, both $XY^\alpha$ and $Z^\alpha$ are clearly non-zero, 
showing a canted $120^\circ$ order on the $A$ sublattice 
and $z$-polarization on the $B$ sublattice. 
Moreover, we establish the existence of QAH effect directly from 
the evolution of the entanglement spectrum as a $h/e$ flux quantum 
is threaded adiabatically through the cylinder~\cite{Zaletel2014,supp}.

In Fig~\ref{fig:DMRGFig}b, we show
the response of the QAH phase to a magnetic Zeeman field, $H_z = -\frac{h}{2}\sum_{i} (n_{i\uparrow} - n_{i\downarrow})$.
The Hall conductivity $\sigma_{xy}$ changes sign abruptly at $h=0$.
There is a discontinuity in $Z$ at $h = 0$ due to broken symmetry, after which total $|Z|$ increases smoothly with $h$.
This is possible via an increase in canting of the $A$ sublattice, and is a signature of our QAH phase with 
a partial spin $s_z$ polarization---as opposed to fully saturated---at zero field. 

We also comment on the stability of the QAH phase against nearest neighbor repulsion, $H_V = V \sum_{\langle i,j\rangle} n_i n_j$.  
At small $V$, the QAH remains present, albeit in a narrower range of $\Delta$~\cite{supp}.
When $V$ is sufficiently large,  
an abrupt transition between $A$- and $B$-sublattice polarized Mott insulators is found around $\Delta=0$, without the intervening QAH phase.     


\begin{figure}[t]
\includegraphics[width=\columnwidth]{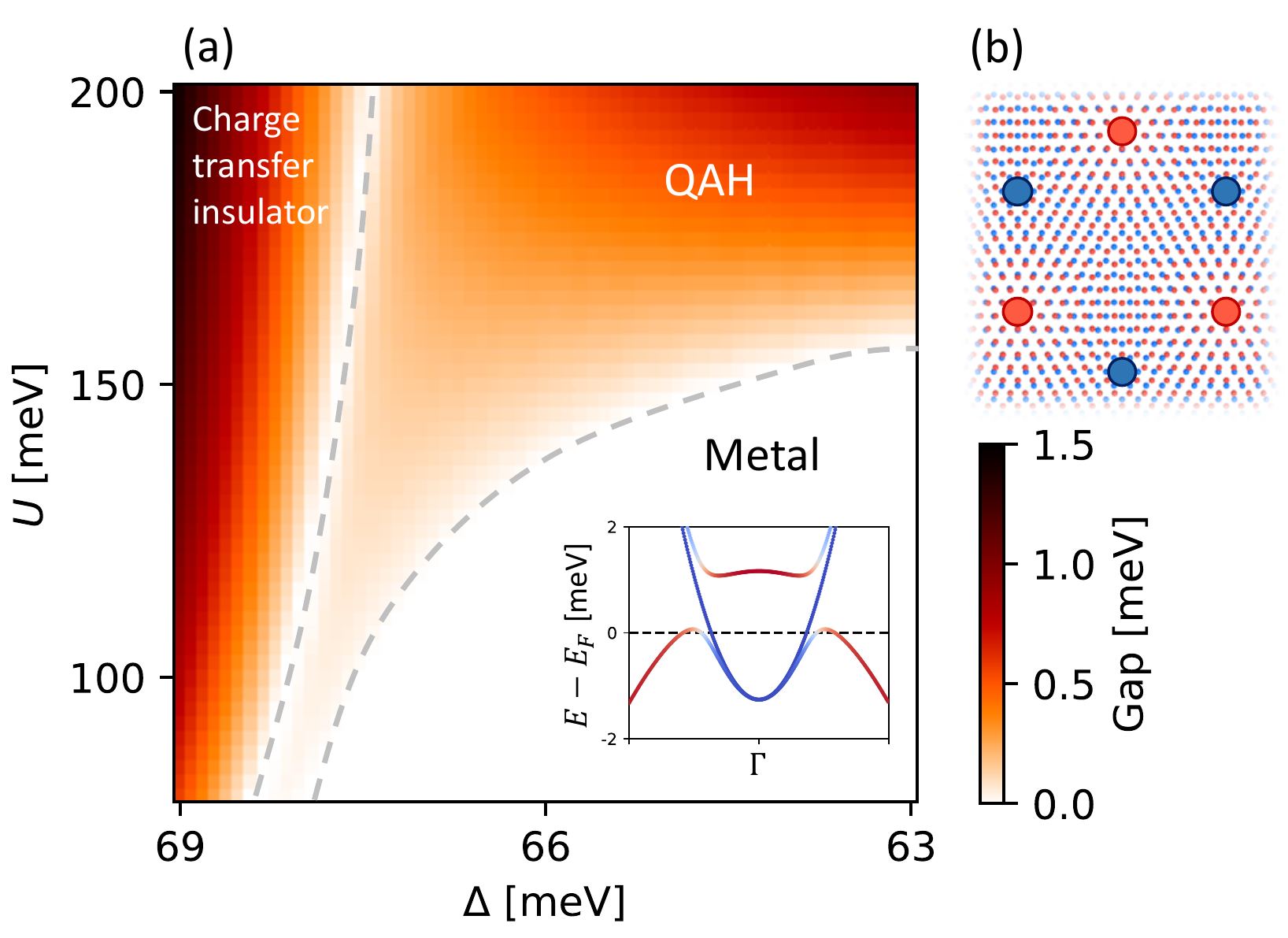}
\caption{
(b) HF phase diagram using the realistic model parameters $(t_A,t_B,t_{AB})=(4.5,9,2)$meV~\cite{supp} describing holes in MoTe$_2$/WSe$_2$.
Color indicates the charge gap.
We find Mott, QAH, and metal phases near band inversion.
Inset shows the quasiparticle bands near the Fermi energy in the metal phase at $\Delta=65$meV and $U=100$meV.
Note that our tight binding model describes holes in this system, hence these bands are minus the electron bands.
(b) Illustration of the moir\'e superlattice in MoTe$_2$/WSe$_2$.  
Low energy hole states on the MoTe$_2$ layer are localized on the MM (red), and WSe$_2$ on the XX (blue) regions. 
Together, they form an effective honeycomb lattice.
}\label{fig:TMDFig}
\end{figure}

Let us now apply our theory to TMD bilayers and in particular, AB-stacked MoTe$_2$/WSe$_2$ heterobilayer. 
Our theory provides a direct explanation for the observed transition from a Mott insulator to a QAH state 
in MoTe$_2$/WSe$_2$ at $n=1$ filling of holes, driven by the applied displacement field~\cite{Tingxin2021}. 
 Our tight binding model captures the topology and essential features of the topmost valence bands from the two layers after a particle-hole transformation.
The role of the displacement field is to decrease the band offset between the two layers, or equivalently, 
reduce the charge transfer energy $\Delta$. For $\Delta$ below a critical value $\Delta_c>0$, 
the quasiparticle gap between MoTe$_2$ and WSe$_2$ Hubbard band is inverted, leading to a QAH insulator.  

Figure~\ref{fig:TMDFig}a shows the HF phase diagram
calculated using realistic parameters for MoTe$_2$/WSe$_2$~\cite{supp}, as a function of $\Delta$ and $U$ near band inversion.  
As $\Delta$ is reduced, we find that the Mott insulating phase transitions into the non-coplanar QAH phase, which 
further transitions into a metal for $U\lesssim 160$meV. 
In this metallic phase, the bands are deeply inverted beyond the $U_B=0$ quadratic band touching regime ($\delta < -4\lambda^2 m_B$ in our effective theory) and $U$ is not large enough to spin polarize the $B$ band. 
The resulting quasiparticle band structure, shown in the inset of Fig~\ref{fig:TMDFig}a, 
feature a nearly spin-degenerate hole pocket on the WSe$_2$ layer and a spin-non-degenerate electron pocket on the MoTe$_2$ layer.  
Thus, this metal phase is a compensated semimetal with $xy$ magnetic order and small quasiparticle Fermi surfaces.
Our phase diagram showing Mott insulator, QAH state, and compensated semimetal as a function of displacement field agrees with the experimentally observed phases in MoTe$_2$/WSe$_2$~\cite{Tingxin2021}. 

Our theory further predicts that (1) at small displacement field, the Mott insulator on the MoTe$_2$ layer is an  
intervalley coherent ($xy$ ordered) state; (2) the QAH state displays {\it partial} valley $z$ polarization on both layers, and simultaneously, intervalley coherence on the MoTe$_2$ layer. 
The $z$ and $xy$ components of the valley order parameter increase and decrease with the displacement field, respectively. 
The spontaneous valley $z$ polarization predicted in the QAH phase (but not in the Mott insulator)  
and its increase with displacement field 
can be detected by magnetic circular dichroism from exciton spin splitting at zero field.  
The existence of intervalley coherence, predicted for both Mott and QAH phase, can be established through 
gapless spin wave transport~\cite{Bi2021}, 
which can be detected by optical means as demonstrated in other TMD heterobilayers~\cite{Jin2018}.

In the lightly inverted regime, our QAH state features a predominantly $xy$ magnetic order, 
with only a small $z$ component. 
It differs from the QAH state in magnetically doped topological insulator films~\cite{Chang2013}, where the magnetic moments spontaneously polarize along $z$ direction. 
Our case should also be contrasted with a fully valley-polarized QAH state that 
arises from topological flat bands with valley-contrasting Chern numbers, as widely discussed for 
magic-angle graphene~\cite{Sharpe2019, Serlin2020, Chen2020, Ming2020, Zhang2019b} and recently proposed for slightly twisted TMD homobilayers~\cite{devakul2021magic}.
 This scenario was also proposed for MoTe$_2$/WSe$_2$~\cite{xie2021theory}.
In the cases discussed above, full valley polarization would be expected throughout the QAH phase.
In contrast, we predict that the spontaneous valley polarization is zero prior to inversion, and develops smoothly in the QAH phase after inversion.
Our work therefore uncovers a general mechanism by which QAH can emerge in the absence of flat bands.





Our mechanism of QAH from inverted Hubbard bands in charge transfer insulators is robust and does not rely on fine tuning.
The effective theory (\ref{eq:effective}), which only involves low-energy quasiparticles, is universally applicable 
in the vicinity of gap inversion, provided that prior to inversion: (1) the charge transfer insulator has a direct quasiparticle band gap; and (2) its electron and hole states at the gap edge have different symmetry eigenvalues.
Note that these requirements are for the quasiparticle band structure of an interaction-induced insulator, {\it not} the noninteracting band structure. 

The central idea of this work, creating magnetic topological states by inverting the charge transfer gap, 
is potentially applicable to a broad range of materials. Besides MoTe$_2$/WSe$_2$, twisted TMD homobilayers under a displacement field also realize a two-band Hubbard model with a tunable charge transfer energy, and therefore may display a similar QAH phase without requiring magic-angle flat bands. Another promising platform is heterostructures between two-dimensional semiconductors and magnetic insulators. We also note the possibility of negative charge transfer gap in transition metal oxides~\cite{Ushakov2011,Choudhury2015}
and 
perovskite nickelates~\cite{Bisogni2016}, which may provide a new venue for topological physics. 

\emph{Acknowledgement} ---
We are grateful to Yang Zhang, Valentin Crepel, Kin Fai Mak, Jie Shan, Shengwei Jiang, and Tingxin Li for helpful discussion on this work and related collaborations.
This work is funded by the Simons Foundation through a Simons Investigator Award.
LF is partly supported by the David and Lucile Packard Foundation.






\bibliographystyle{unsrt}
\bibliography{ref}

\clearpage
\newpage
\appendix

\section{Details of numerical calculation}
\subsection{Hartree-Fock}
In this Appendix, we present more details on our HF calculations.  
We employ two different approaches. 
In the first, we perform the self-consistent HF calculation using a 6 site unit cell commensurate with the expected $\sqrt{3}\times\sqrt{3}$ order.  
In the second, we use $U_{\zeta}$ to transform the Hamiltonian and perform self-consistent HF assuming translation invariance (a 2 site unit cell), and pick the one $\zeta=0,\pm$ with lowest energy.  
These transforms any potential $xy$ AFM order into $xy$ FM order.

The advantage of the second approach is that it does not reduce the BZ and is conceptually simpler, with only a single filled quasiparticle band, whereas in the first approach one must work in a reduced BZ with three filled bands.
The band structure and Berry curvature in Fig 2 (c,d) of the main text are computed from this second approach.

The disadvantage of the second approach is that it is not able to capture all types of spin orders.
For example, it cannot describe a state with different wavevectors on the two sublattices, such as a state with $120^\circ$ $xy$ AFM on the $A$ sublattice and $xy$ FM on the $B$ sublattice.
However, in the range of parameters we have examined, we find that these two approaches converge to the same result, indicating that such spin configurations do not appear.

We obtain the self-consistent HF solution by iteration.  
We consider an initial starting values for density $\langle n_i\rangle$ and spin $\langle\vec{s}_i\rangle$ expectation values.
In the first approach, we start with a $z$ FM, $xy$ FM, $\zeta=\pm1$ $xy$ AFMs, and in the second approach, we consider $z$ and $xy$ FM phases, all of which are sublattice balanced $n_A=n_B=\frac{1}{2}$.
In addition, we add a small random noise of order $\sim 0.01$ to the initial starting expectation values.
Using these expectation values, the Hubbard term in the Hamiltonian is then replaced by 
\begin{equation}
H_{\rm{Hub}}^{\rm{HF}} = \frac{U}{2} \sum_{i} \left(n_{i} \langle n_{i} \rangle - \vec{s}_{i}\cdot \langle\vec{s}_{i}\rangle -\frac{1}{2}\langle n_{i}\rangle^2 + \frac{1}{2}|\langle\vec{s}_{i}\rangle|^2\right)
\end{equation}
and diagonalized, in an $N_k\times N_k$ momentum space grid.
We use $N_k=180$.
The new expectation values $\langle n_i \rangle^\prime$,$\langle\vec{s}_i\rangle^\prime$, are then calculated at filling $n=1$.  
The calculation is then repeated using these new expectation values to construct the HF Hamiltonian.
We repeat the calculation until the norm of the difference between consecutive iterations, $D = \sqrt{\sum_i |\langle n_i\rangle^\prime-\langle n_i\rangle|^2 + |\langle \vec{s}_i\rangle^\prime-\langle \vec{s}_i\rangle|^2}$, falls below the threshold $D\leq 10^{-10}$.
In the case when different initial values converge to different states, the one with the lowest energy is chosen.

\begin{figure*}
\includegraphics[width=\textwidth]{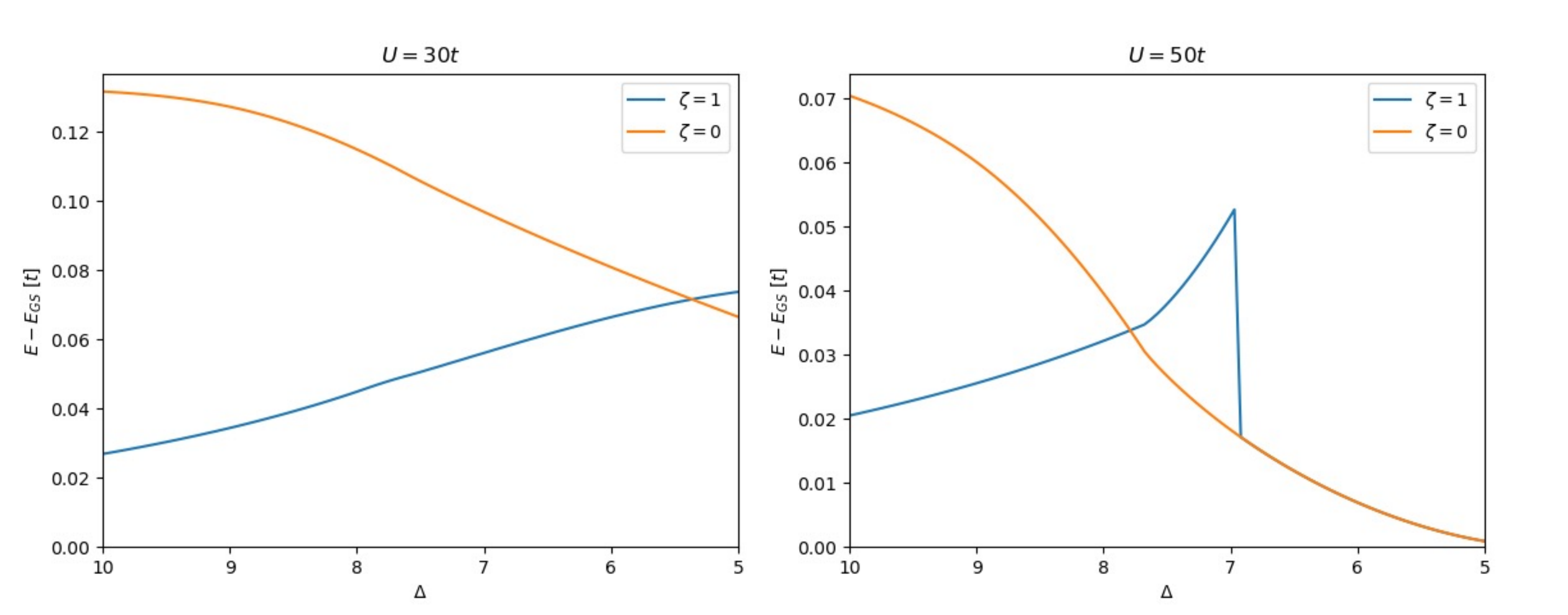}
\caption{
 Plots of the energy of the self-consistent Hartree-Fock solutions with $\zeta=0,+1$, compared to the ground state energy $E_{\rm{GS}}$, which has $\zeta=-1$.
}
\label{appfig:HFE}
\end{figure*}
Throughout the parameter range shown in Fig~\ref{fig:HFFig}a,b, the ground state can be well captured by a state with $\zeta=-1$.  
We may directly compare the energy of this state with the states of the Hamiltonian restricted to $\zeta=0,+1$, by transforming the Hamiltonian using $U_\zeta$ and obtaining the lowest energy self-consistent HF solution with enforced translation invariance.  
The energy differences are shown in Fig~\ref{appfig:HFE} for $U=30t$ and $U=50t$.
As can be seen, the state with $\zeta=-1$ is always lowest in energy throughout the entire range.  

For $U=50t$, as $\Delta$ is decreased below $5t$, a fully $Z$ polarized state becomes favored in HF.  
This is an artifact of the HF method: HF captures the energy of the fully polarized state \emph{exactly} (as the Hubbard interaction energy is identically zero for the fully polarized state), while it merely provides an upper bound for the energy of the strongly correlated AFM state.
Thus, HF overestimates the favorability of the fully polarized state.
This is corroborated by the fact that DMRG does not observe any tendency towards the fully polarized state within this parameter regime.

\subsection{DMRG}
In this Appendix, we present a more detailed description of our numerical DMRG calculation.  
As stated in the main text, we employ the infinite DMRG (IDMRG) algorithm on an infinite cylinder.
We take the XC geometry, in which one of the nearest neighbor bonds is oriented in the $x$ (infinite) direction.
We utilize a $1\times L_y$ unit cell with boundary conditions commensurate with the $\sqrt{3}\times\sqrt{3}$ order.
The sites are ordered in DMRG starting with $A$ sublattice sites in order of increasing $y$, and then again for $B$ sublattice.
We begin with a random product state of fermions in the $S^z$ basis, at the desired density of $n=1$ fermions per unit cell, and total $S^z=0$.

The IDMRG algorithm is performed with conserved quantum numbers corresponding to total particle number $N=N_{\uparrow} + N_{\downarrow}$, and spin $S^z$ parity, $(-1)^{N_{\uparrow}-N_{\downarrow}}$.
Although the Hamiltonian has spin-$U(1)$ symmetry, and therefore conserved total $S^z$, we choose to only conserve the parity.
The reason for this is so that states which spontaneously break spin-$U(1)$ symmetry, such as $xy$ ordered states, can be represented and diagnosed explicitly.
Thus,  $xy$ long-range ordered states can be diagnosed simply via non-zero expectation value $\langle S^{x,y}\rangle \neq 0$, rather than spin-spin correlation functions.
Furthermore, this allows us to access states with $S^z$ density does not correspond to a particular choice of $N_{\uparrow},N_{\downarrow}$, in the $1\times L_y$ unit cell.
This is important as the non-coplanar QAH phase has smoothly varying $S^z$ as a function of $\Delta$.

We also consider applying flux $\psi$ through the cylinder.
This is modeled by modifying the hopping terms such that a fermion picks up an additional phase factor $e^{i\psi}$ upon going around the circumference of the cylinder.
We compute the order parameters in Figure~\ref{fig:DMRGFig} for fluxes $\psi=0,\pi$.
In the 2D limit, $L_y\rightarrow\infty$, all observables should be independent of $\psi$.
Thus, the difference of observables between $\psi=0$ and $\psi=\pi$ is an indication of finite circumference effects.
In Figure~\ref{fig:DMRGFig} of the main text, we plot the average of the order parameters obtained for $\psi=0,\pi$, and the error bar indicates the difference, on a $L_y=6$ cylinder with maximum bond dimension $\chi=1600$.

\begin{figure*}[t]
\includegraphics[width=\textwidth]{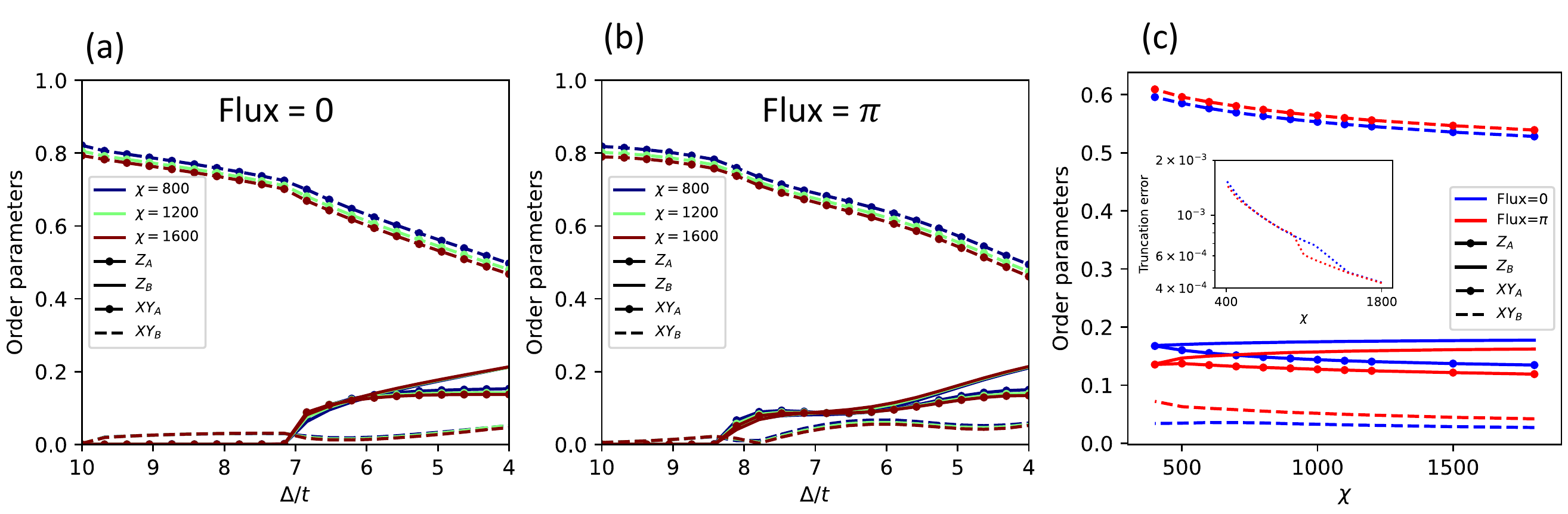}
\caption{
Plot of the ordered parameters as a function of $\Delta$ obtained from DMRG for (a) zero and (b) $\pi$ flux threaded through the cylinder.
In (c), we fix $\Delta=5t$ and show the order parameters as a function of bond dimension $\chi$.  
Inset shows the DMRG truncation error.
}
\label{appfig:DMRG}
\end{figure*}
\begin{figure}[t]
\includegraphics[width=\columnwidth]{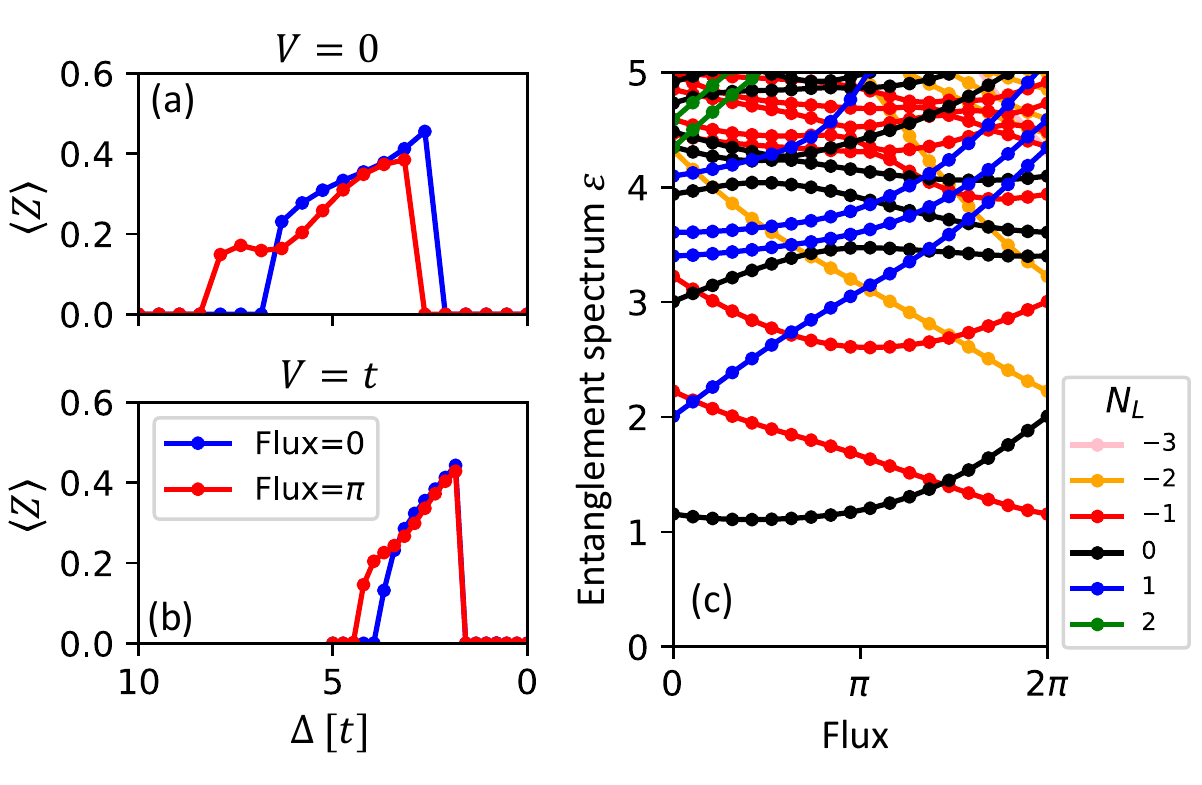}
\caption{
The total $Z=Z_A+Z_B$ as a function $\Delta$ is shown for nearest neighbor repulsion (a) $V=0$ and (b) $V=t$.
The physics is qualitatively similar, except occuring within a smaller range of $\Delta$.
In (c), we verify the non-trivial topology of this phase in DMRG by computing the particle number resolved entanglement spectrum adiabatically as $2\pi$ flux is threaded through the cylinder.
We use a maximum bond dimension $\chi=800$.
}
\label{appfig:VES}
\end{figure}

In Figure~\ref{appfig:DMRG}a,b, we show the order parameters for $\psi=0,\pi$ as a function of $\Delta$, for bond dimensions $\chi=800,1200,1600$.
Unless stated otherwise, we use the parameters $t_A=t_B=\frac{1}{2} t_{AB} \equiv t$ and $U_A=U_B=U=50t$.
As can be seen, there is only a small difference in the value of the order parameters as $\chi$ is increased.
In Figure~\ref{appfig:DMRG}c, we focus on $\Delta=5t$, and the dependence of various quantities on bond dimensions from $\chi=400-1800$ is shown.
Importantly, we find that IDMRG converges to a state with canted AFM order: finite $\langle S^z\rangle$ and $\langle S^{x,y}\rangle$ in the $120^\circ$ configuration with $\zeta=-1$, as defined in the main text.

In Figure~\ref{appfig:VES}a,b, we show the effect of nearest neighbor repulsion $V$.  
The QAH phase, identified by a non-zero $Z=Z_A+Z_B$ expectation value, persists in a finite window of $\Delta$.
The leading effect of a small $V$ is to narrow the range in which this phase appears. 
For large $V$, there is a first order transition directly from an $A$-sublattice polarized Mott state to a $B$-sublattice polarized state.

To confirm the non-trivial topology of this phase, we examine the entanglement spectrum  under an adiabatic threading of $2\pi$ flux through the cylinder~\cite{Zaletel2014,Gong2014,Zhu2016,Laughlin1981}.
Considering a cut in the cylinder, the ground state computed in IDMRG is naturally represented by the a Schmidt decomposition $\ket{\psi} = \sum_{i} \lambda_{i} \ket{i_L}\otimes \ket{i_R}$, where $\ket{i_{L(R)}}$ forms an orthonormal basis for states on the left (right) side of the cut, and $\lambda_{i}^2\equiv e^{-\varepsilon_{i}} >0$ is the ES.
As we explicitly conserve particle number, each $\varepsilon_i$ can be labeled by the integer particle number $N_{L,i}$ associated with the state left state $\ket{i_L}$.
In Figure~\ref{appfig:VES}c, we show the evolution of the entanglement spectrum $\varepsilon_{i}$ in the ground state at $\Delta=5t$ as flux is threaded through the cylinder, with $N_{L,i}$ indicated by color.
As can be seen, the spectrum comes back to itself after $\psi=2\pi$ flux is threaded, however, the associated particle numbers $N_{L,i}$ only comes back to itself minus one.
This indicates that, upon threading $2\pi$ flux, one particle of charge is pumped from the left of the cut to the right.  
This charge pumping is direct proof of the non-trivial Chern number and QAH effect in the ground state.

We comment on the small but non-zero $XY_B$ order parameter in the QAH phase, which is found in DMRG but is absent in HF.  
This corresponds to a small $XY$ component on a $B$ site which is aligned with the $XY$ component of one of its neighboring $A$ sites along the length of the cylinder.
Taken seriously, this would correspond to a nematic order when extrapolated to the infinite honeycomb lattice.
However, we believe this is likely an artifact arising from the intrinsic anisotropy of the cylindrical geometry used in DMRG.
The geometry explicitly breaks rotation symmetry, which may explain the small nematic component.
Additional numerical work for larger cylinder circumferences is necessary to ascertain the source of this apparent nematicity.

In the magnetic field calculation in Fig~\ref{fig:DMRGFig}b of the main text, each data point represents an independent DMRG calculation.  Thus, we do not see hysteresis, which would be expected if one slowly swept $h$. 

We have also performed IDMRG calculations with boundary conditions commensurate with a $2\times 2$ unit cell (which can capture tetrahedral or stripe magnetic order, for example).  
The resulting state attempts to form the $120^\circ$ order, but is unable due to incommensuration with the IDMRG unit cell. 
The resulting energy is higher than that of the $\sqrt{3}\times\sqrt{3}$ commensurate phase.

\section{Relation to moir\'e TMD bilayers}
In this Appendix, we discuss in detail the connection between our tight binding model and moir\'e TMD bilayers, and especially the origin of the phase factors $\phi_\alpha$ mentioned in the main text.
Specifically, our tight binding model captures the main qualitative features of the first bands of each layer and valley in $K$-valley derived moir\'e TMD bilayer systems with an effective honeycomb lattice description.
As we shall show, the two sublattices correspond to the two layers, and the spin corresponds to the valley $\pm K$ degrees of freedom.

To motivate the tight binding model, we begin with a single layer $\alpha$, in which the low energy degrees of freedom are spin-polarized at the $\pm K_\alpha$ points of the original BZ.  
In the presence of a second layer, effects such as lattice relaxation result in an effective potential in the first layer.  
The moir\'e bands can be well described by a continuum model description~\cite{Wu2019,Pan2020}.
Neglecting interlayer tunneling for now, the $\alpha$ layer Hamiltonian can be well described by an effective mass description of the electron (or hole) dispersion about the $\pm K_\alpha$ points in the presence of a periodic potential with the moir\'e period, \begin{equation}
H^{\mathrm{cont}}_\alpha = \frac{(\bm{k}-\sigma \bm{K}_\alpha)^2}{2m_\alpha} + V_\alpha(\bm{r}) ,\label{eq:Hcont}
\end{equation}
 where $\sigma=\pm $ encodes the $\pm K$ valley degree of freedom.
Without fine-tuning, the potential $V_\alpha(\bm{r})$ will generically have a minimum at one of the high symmetry stacking regions of the moir\'e structure, forming a triangular superlattice.
The resulting bands in the reduced moir\'e BZ can be described by a triangular lattice tight binding model in terms of localized Wannier orbitals centered at these potential minima with valley pseudo-spin internal degree of freedom.
The Hamiltonian $H_\alpha$ in the main text contains only the nearest neighbor hopping term.
Incorporating the triangular lattices of both layers then results in an effective honeycomb lattice, as long as the potential minima of the two layers lie at different high symmetry positions in the moir\'e unit cell.  
The interlayer tunneling gives rise to the hopping term $H_{AB}$.  
The effective tight binding model incorporates the first band of each valley and layer, which is sufficient in describing the physics at filling $n=1$ which do not involve any higher bands.
 Additional terms arising from neglected higher bands or strain~\cite{xie2021theory,zhai2020theory,magorrian2021multifaceted,tang2021geometric} may be present, although large-scale DFT on fully relaxed structure indicates that such terms are small compared to the potential term~\cite{ZhangPNAS}.

In order to discuss the finer details of the mapping to the tight binding model,
 we must first address two things:
the folding to the moir\'e BZ and the $C_{3}$ eigenvalues at high symmetry momenta.

First, we discuss the folding: specifically where the points $K_A$ and $K_B$, which determines the position of the band minimum, fold to in the moir\'e BZ.  
In general, the true folding will depend on the precise commensurate structure of the bilayer.  
For example, the moir\'e structure of MoTe$_2$/WSe$_2$ is close to the commensurate approximation of $13\times 13$ MoTe$_2$ ($A$) unit cells and $14\times14$ WSe$_2$ ($B$) unit cells~\cite{ZhangPNAS}.  
In this case, the folding to the moir\'e BZ is $K_A (= 13K)\cong K$ and $K_B\cong -K$.
On the other hand, another close approximation is $14\times 14$ $A$ and $15\times 15$ $B$ unit cells, in which case $K_A\cong -K$ and $K_B\cong \Gamma$.
However, the precise folding should not affect any physical observables on the moir\'e scale (since, in general, the structure need not even be commensurate).
In this sense, there is a freedom of choice in selecting a folding scheme.
We define folding schemes by $\xi=0,\pm 1$, such that $K_A\cong (\xi+1)K$ and $K_B\cong(\xi-1)K$.
The two folding schemes mentioned above for MoTe$_2$/WSe$_2$ correspond to $\xi=0$ and $\xi=1$, respectively.

Second, we discuss the $C_{3}$ eigenvalues.  We define $C_{3}$ to be a $2\pi/3$ counter-clockwise rotation about the $z$ axis centered at the MM region where two metal atoms from both layers lie on top of each other.
The $C_{3}$ eigenvalues are determined by the wavefunction of the monolayer at $K$, and the position of the Wannier center.
Let us denote the $\alpha$ layer monolayer $C_{3}$ eigenvalue at $\sigma K$ as $e^{\frac{2\pi i}{3} \sigma j_\alpha}$ ($j_\alpha$ is half integer due to spin-$\frac{1}{2}$).
For the TMD heterobilayer, we label the three high symmetry stacking positions in a moir\'e unit cell as $\bm{R}_n = \frac{a_M}{\sqrt{3}}(0,n)$ for $n=0,1,2$, corresponding to MM, XX, and MX stacking regions respectively.
For the folding scheme $\xi$, the $C_{3}$ eigenvalues of the first band of valley $\sigma$ and layer $\alpha$ at momentum $\ell K$ ($\ell=0,\pm1$) is given by
\begin{equation}
\Theta_{\alpha}^{\sigma}(\ell K)=\exp\left(\frac{2\pi i}{3}[\sigma j_\alpha + (\ell - \sigma[\xi + \tau_\alpha]) n_\alpha]\right)
\label{eq:Theta}
\end{equation}
where $n_\alpha$ is such that $\bm{R}_{n_\alpha}$ is the position of the Wannier centers, and $\tau_A=-\tau_B=1$.

The role played by the $C_3$ eigenvalues is crucial: the interlayer tunneling only couples states at the same high-symmetry momentum if they have the same $C_3$ eigenvalue.
Our tight binding model describes the folding choice $\xi_0$ in which the $C_3$ eigenvalues match at $\Gamma$, 
\begin{equation}
\xi_0 = (n_B-n_A)(j_B-j_A + n_A+n_B)\mod 3 .
\end{equation}
Note that $n_B-n_A\neq 0$ since we assume a honeycomb lattice structure.
Other folding choices can be described by the shifted Hamiltonians, $U_\zeta\mathcal{H}U_{\zeta}^{\dagger}$, in which the nearest neighbor hoppings are direction and spin-dependent.

Next, the phase factors $\phi_\alpha$ in the tight binding model should be chosen to describe the correct band dispersion.
If there had been no momentum offset in Eq~\ref{eq:Hcont}, then we would have had $\phi=0$.  
A momentum shift to $\sigma K_\alpha \cong \sigma(\xi+\tau_\alpha)K$ corresponds to $\phi_\alpha = \frac{2\pi}{3}(1+\tau_\alpha\xi)$.

We also note that in TMD bilayers, $\phi_\alpha$ is not strictly fixed to be a multiple of $2\pi/3$.  
The above analysis gives an estimate of $\phi_\alpha$ in order to match the topology and positions of the maxima and minima.  
However, small deviation of $\phi_\alpha$ from this value is expected in real systems (see, for example, Ref~\cite{devakul2021magic}).

Let us take AB-stacked MoTe$_2$/WSe$_2$ as an example. 
For the MoTe$_2$ layer, the $C_3$ eigenvalue of the state at $K$ is $e^{-\frac{i\pi}{3}}$ and
the moir\'e bands localized at the MM region~\cite{ZhangPNAS}: thus $j_A=-\frac{1}{2}$ and $n_A=0$.  
Similarly for the WSe$_2$ layer, due to AB-stacking we have $j_B=\frac{1}{2}$, and the XX localized wavefunction corresponds to $n_B=1$.
Direct calculation of $C_3$ eigenvalues from large-scale DFT is in agreement with Eq~\ref{eq:Theta}, with the folding choice $\xi=0$~\cite{ZhangPNAS} (note our definition of $K$ is opposite to that of Ref~\cite{ZhangPNAS}).
Our tight binding model describes the folding $\xi_0=-1$, in which $K_A\cong \Gamma$ and $K_B\cong K$.
This corresponds to the phase parameters $\phi_A=0$ and $\phi_B = -\frac{2\pi}{3}$ (mod $2\pi$), as used in the main text.
In homobilayer systems, such as small angle twisted WSe$_2$/WSe$_2$, as long as there is a honeycomb lattice description at small angles, $\xi_0=0$ is fixed by symmetry and we have $\phi_\alpha\approx \frac{2\pi}{3}$.
Tight binding models for other folding choices $\xi\neq\xi_0$ are described by the shifted Hamiltonians $U_\zeta \mathcal{H} U_{\zeta}^\dagger$ with $\zeta = \xi-\xi_0$.

For either of $\xi_0=0,-1$, the band inversion at positive $\Delta$ is topological and the physics discussed in the main text applies.  
For $\xi_0=+1$, the $C_3$ eigenvalues match at band inversion, and there is no topological band inversion.  
There is another band inversion starting from the fully occupied $B$ sublattice at $\Delta \rightarrow-\infty$ and reducing $|\Delta|$, which is topological for $\xi_0=0,+1$, but non-topological for $\xi_0=-1$.

\begin{figure}[t]
\includegraphics[width=\columnwidth]{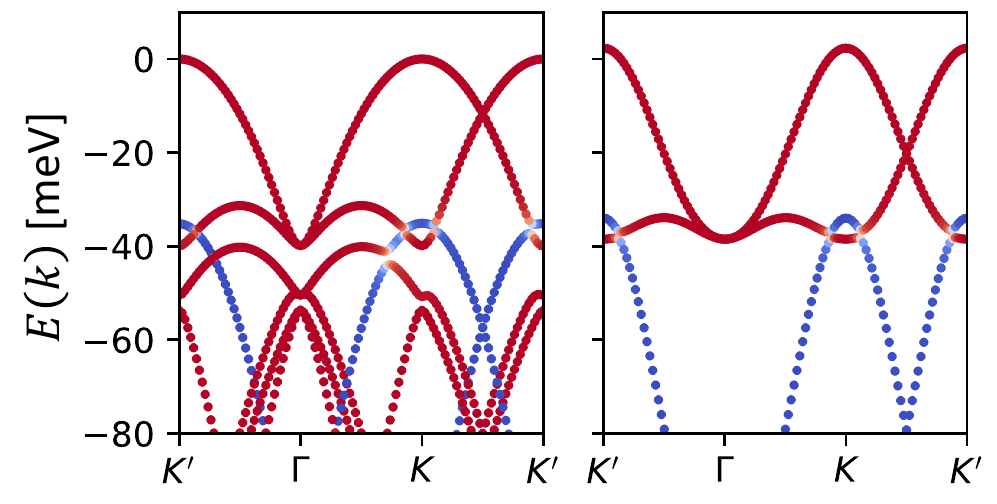}
\caption{(left) The continuum model noninteracting bands for AB-stacked MoTe$_2$/WSe$_2$ valence bands~\cite{ZhangPNAS} and  (right) the tight binding model approximation is shown, near noninteracting band inversion.
The tight binding model captures the qualitative features of the first bands of each valley and layer.  
Color indicates layer content: MoTe$_2$ is red, and WSe$_2$ is blue.
}\label{appfig:TBCont}
\end{figure}

The magnitude of the hopping terms, $t_A$, $t_{AB}$, and $t_{AB}$, can be fit to best match the band structure from large-scale DFT.  
In Figure~\ref{appfig:TBCont}, we show the continuum model bands for MoTe$_2$/WSe$_2$ with parameters from Ref~\cite{ZhangPNAS} and the tight binding model bands  with parameters $t_A=4.5$meV, $t_B=9$meV, and $t_{AB}=2$meV, which roughly matches the band widths and the magnitude of the interlayer tunneling.
The bands of $-\mathcal{H}$ is plotted (due to particle-hole transformation) and the folding choice $\xi=0$ is used, to conform with Ref~\cite{ZhangPNAS}.
To match the exact shape of the band requires further range hoppings, but should not qualitatively affect the universal physics near band inversion as described in the main text.

Finally, let us briefly discuss Ref~\cite{xie2021theory}, which presents an alternate explanation for the QAH phase in MoTe$_2$/WSe$_2$.  
The mechanism for topology suggested by Ref~\cite{xie2021theory} is that a strain-induced pseudomagnetic field (a term left out of Eq~\ref{eq:Hcont}) may cause the noninteracting first moir\'e band of the MoTe$_2$ layer to carry non-trivial valley-contrasting Chern number. 
Interactions then induce a fully valley polarized state, resulting in QAH.
However, the strain-induced topology is not supported by fully-relaxed large-scale DFT~\cite{ZhangPNAS}, which shows topologically trivial first MoTe$_2$ bands.
Also, the self-consistent HF calculation in Ref~\cite{xie2021theory}, used to argue for a fully valley polarized state, assumes translation invariance and therefore misses the $120^\circ$ ordered state which we find is significantly more energetically favorable.

\section{Additional details of HF bands}

\begin{figure*}[t]
\includegraphics[width=\textwidth]{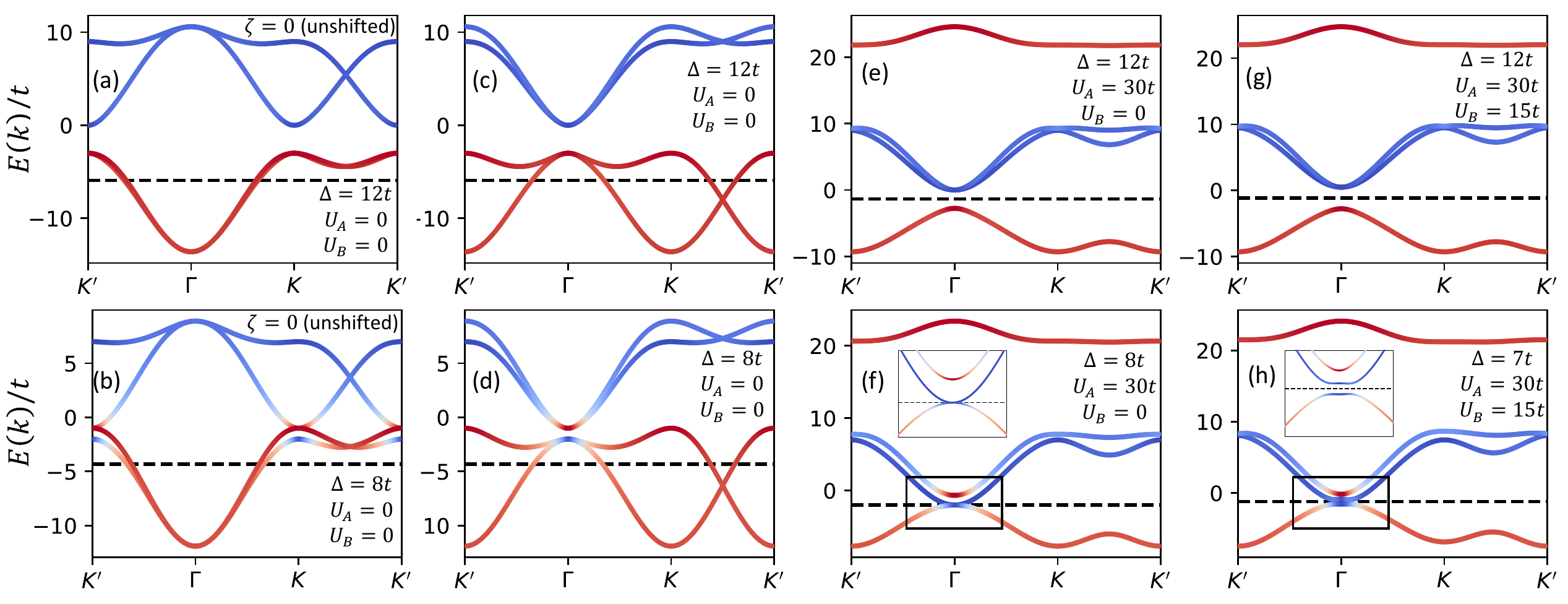}
\caption{Plot of the HF band structure (see accompanying discussion) as various interactions $U_A,U_B$ is included, before and after band inversion.
Black dashed line indicates the Fermi energy at filling $n=1$.
Color indicates the sublattice content: $A$ sublattice in red, and $B$ in blue.
} \label{appfig:HFUab}
\end{figure*}

In this Appendix, we discuss additional details and HF band structures related to the discussion in the main text.
We use a representative set of parameters $t_A=t_B=\frac{1}{2} t_{AB} \equiv t$ and $\phi_A=0,\phi_B=-\frac{2\pi}{3}$, as in the main text.
Fig~\ref{appfig:HFUab} shows the HF band structure at various interactions $U_A,U_B$, and charge transfer energies $\Delta$.

First, in Fig~\ref{appfig:HFUab}(a,b), we show the noninteracting band structure ($U_A=U_B=0$) of $\mathcal{H}$, prior to the spin-dependent shift $U_\zeta$, before and after band inversion.  

Next, we apply the shift $U_\zeta$.  
The noninteracting band structure of the shifted Hamiltonian, $\mathcal{H}^\zeta = U_\zeta \mathcal{H} U_\zeta^\dagger$, is given by
\begin{equation}
H_\sigma^\zeta(\bm{k}) = \begin{pmatrix}
\mathcal{E}^{\zeta}_{A\sigma}(\bm{k}) & T^\zeta_\sigma(\bm{k}) \\
T^\dagger_\sigma(\bm{k}) & \mathcal{E}_{B\sigma}(\bm{k})
\end{pmatrix}
\end{equation}
for spin $\sigma$, where
\begin{equation}
\mathcal{E}_{\alpha\sigma}^\zeta(\bm{k}) = -2 t_\alpha\sum_n\cos\left(\bm{k}\cdot \bm{a}_n + \frac{2\pi\zeta s_\sigma}{3} +  s_\sigma\tau_\alpha\phi_\alpha\right) - \frac{1}{2}\tau_\alpha\Delta
\end{equation}
\begin{equation}
T_{\sigma}^\zeta(\bm{k}) = - t_{AB} \left( e^{\frac{2\pi i \zeta s_\sigma}{3}}e^{-i \bm{k}\cdot\bm{b}_1} + e^{-\frac{2\pi i \zeta s_\sigma}{3}}e^{-i \bm{k}\cdot\bm{b}_2} + e^{-i\bm{k}\cdot\bm{b}_3}\right)
\end{equation}
Notice that the nearest neighbor hopping term of the shifted model, $T^\zeta_\sigma$, is now both spin and direction dependent.
The bands in Fig~\ref{appfig:HFUab}(c-h) are plotted with the shift $\zeta=-1$, which transforms the observed A sublattice $xy$ AFM to an $xy$ FM.

First, in Fig~\ref{appfig:HFUab}(c,d), we show the noninteracting band structure from Fig~\ref{appfig:HFUab}(a,b), but with the shift $U_{\zeta=-1}$ applied.  
This shifts both band minima of the B sublattice bands to $\Gamma$.
Without interactions, the ground state is a metal with partial filling of both $\sigma$ bands on the A sublattice.  

Next, Fig~\ref{appfig:HFUab}(e,f), shows the HF band structure with $U_A=30t$ and $U_B=0$, before and after band inversion.
The A bands (red) is split into lower and upper Hubbard bands, separated by a Mott gap $\sim U$.  
Before band inversion, the ground state is insulating with full filling of the lower Hubbard band.
Fig~\ref{appfig:HFUab}e shows $\Delta=12t$, where the B bands lie in between the lower and upper Hubbard bands resulting in a charge transfer insulator.
As $\Delta$ is reduced, the charge transfer gap becomes negative.  
Right after band inversion, as shown in Fig~\ref{appfig:HFUab}f, the resulting band structure exhibits a quadratic band touching at the Fermi energy (inset shows magnified band structure).

Finally, Fig~\ref{appfig:HFUab}(g,h) shows the effect of interactions $U_{B}=15t$ on the $B$ sublattice, before and after inversion.
Prior to inversion, $U_B$ does not play an important role due to small density on the $B$ sublattice.
After inversion, the $B$ sublattice develops a spontaneous $z$ polarization due to non-zero $U_B$.
The resulting filled lower Hubbard band, after inversion, carries non-trivial Chern number $\mathcal{C}=\pm 1$.
We remark that other effects, such as an applied Zeeman field in the $z$ direction, can also induce polarization in the $B$ sublattice resulting in Chern bands even at $U_B=0$.

We end with a discussion on the Mott state at large $\Delta$.  
In our tight binding model, $\phi_A=0$ is special in that there is an emergent $SU(2)$ symmetry in the limit $\Delta\rightarrow \infty$ when the $B$ sublattice can be ignored.
The two $xy$ AFMs with $\zeta=\pm$ are degenerate in this limit, along with any in-plane ordered phases beyond $xy$.
This symmetry can be broken in two ways: (1) as $\Delta$ is reduced, the effect of the $B$ sublattice with $\phi_B= \pm \frac{2\pi}{3}$ favors the $xy$ AFM with $\zeta=\pm$, or (2) it may be that $\phi_A$ is non-zero but small, which will favor the $xy$ AFM with $\zeta=-\mathrm{sign}(\phi_A)$.
An interesting scenario arises when these effects favor states with different $\zeta$: for example, $\phi_B=-\frac{2\pi}{3}$ and $\phi_A = -\epsilon$ for small $\epsilon>0$.
For large $\Delta$, the ground state is an $xy$ AFM with $\zeta=+$, due to the sign of $\phi_A$.  
For this state, the quasiparticle gap is indirect and there is no QAH phase.
However, as $\Delta$ is reduced, there is at some point a first-order phase transition to the $xy$ AFM with $\zeta=-$ due to the coupling to the $B$ layer, for which our mechanism for QAH is possible.
Exactly where this transition occurs depends on non-universal details such as the values of the hoppings and $\epsilon$. 
Thus, even though the quasiparticle gap at $\Delta\rightarrow\infty$ is indirect, QAH may still be possible by our mechanism through a transition to a competing ordered state (in which the quasiparticle gap is direct) as $\Delta$ is reduced.

\end{document}